\documentclass[12pt]{article}
\usepackage[dvipdfmx]{graphicx}
\usepackage{amsmath,amssymb,amsfonts,graphicx,slashed,amsthm,mathtools,upgreek, enumerate, tensor, subfig,epsf}
\usepackage{amsmath,braket}
\usepackage{graphicx,hyperref,color}
\hypersetup{
    bookmarks=true,         
    unicode=false,          
    pdftoolbar=true,        
    pdfmenubar=true,        
    pdffitwindow=false,     
    pdfstartview={FitH},    
    pdfnewwindow=true,      
    colorlinks=true,       
    linkcolor=blue,          
    citecolor=blue,        
    filecolor=blue,      
    urlcolor=blue           
}
\newcommand{\be}{\begin{equation}}
\newcommand{\ba}{\begin{eqnarray}}
\newcommand{\ea}{\end{eqnarray}}
\newcommand{\ee}{\end{equation}}

\newcommand{\f}{\frac}
\newcommand{\s}{\sqrt}

\newcommand{\no}{\nonumber \\}
\newcommand{\la}{\langle}
\newcommand{\lb}{\rangle}
\newcommand{\bea}{\begin{eqnarray}}
\newcommand{\eea}{\end{eqnarray}}
\newcommand{\bes}{\begin{equation*}}
\newcommand{\beas}{\begin{eqnarray*}}
\newcommand{\eeas}{\end{eqnarray*}}
\newcommand{\bas}{\begin{array*}}
\newcommand{\eas}{\end{array*}}
\newcommand{\ees}{\end{equation*}}

\newcommand{\p}{\partial}
\newcommand{\ep}{\epsilon}

\newcommand{\ov}{\overline}

\begin{document}

\begin{titlepage}
\thispagestyle{empty}

\begin{flushright}
YITP-21-19
\\
IPMU21-0017
\\
\end{flushright}

\bigskip

\begin{center}
\noindent{{\large \textbf{Spectrum of End of the World Branes in Holographic BCFTs}}}\\
\vspace{1.5cm}

Masamichi Miyaji$^{a}$, Tadashi Takayanagi$^{b,c,d}$ and Tomonori Ugajin$^{b,e}$
\vspace{0.5cm}\\

{\it $^a$ Berkeley Center for Theoretical Physics, \\
Department of Physics, University of California, Berkeley, 
CA 94720, USA }\\
\vspace{1mm}
{\it $^b$Center for Gravitational Physics,\\
Yukawa Institute for Theoretical Physics,
Kyoto University, \\
Kitashirakawa Oiwakecho, Sakyo-ku, Kyoto 606-8502, Japan}\\
\vspace{1mm}
{\it $^c$Inamori Research Institute for Science,\\
620 Suiginya-cho, Shimogyo-ku,
Kyoto 600-8411 Japan}\\
\vspace{1mm}
{\it $^{d}$Kavli Institute for the Physics and Mathematics
 of the Universe (WPI),\\
University of Tokyo, Kashiwa, Chiba 277-8582, Japan}\\
\vspace{1mm}
{\it $^{e}$The Hakubi Center for Advanced Research, Kyoto University,
Yoshida Ushinomiyacho, Sakyo-ku, Kyoto 606-8501, Japan}\\

\end{center}

\begin{abstract}
We study overlaps between two regularized boundary states in conformal field theories.  
Regularized boundary states are dual to end of the world branes in an AdS black hole via the AdS/BCFT. Thus they can be regarded as microstates of a single sided black hole. Owing to the open-closed duality, such an overlap  between two different  regularized boundary states   is  exponentially 
suppressed as $\langle \psi_{a} | \psi_{b} \rangle \sim e^{-O(h^{(min)}_{ab})}$, where $h^{(min)}_{ab}$ is the lowest energy of open strings which connect two different boundaries $a$ and $b$. 
Our gravity dual analysis leads to
$h^{(min)}_{ab} = c/24$ for a pure AdS$_3$ gravity. This shows that a holographic boundary state is a random vector among all left-right symmetric states, whose number is given by 
a square root of the number of  all black hole microstates. 
We also perform a similar computation in higher dimensions, and find that  $h^{( min)}_{ab}$ depends on the tensions of the branes. In our analysis of holographic boundary states, the off diagonal 
elements of the inner products can be computed directly from on-shell gravity actions, 
as opposed to earlier calculations of inner products of microstates in two dimensional gravity.
\end{abstract}

\end{titlepage}

\newpage


\section{Introduction}

 A conformal field theory (CFT) which is dual to classical gravity on an anti de-Sitter space (AdS) via the AdS/CFT \cite{Ma}, is expected to be strongly interacting and to have large degrees of freedom. Such CFTs, called holographic CFTs, have been considered to describe quantum systems 
 with maximal quantum chaos \cite{Maldacena:2015waa}.
 Conformal bootstrap studies predict the special feature that a holographic CFT has a large spectrum gap and that its low energy spectrum is sparse \cite{Heemskerk:2009pn,Hartman:2014oaa,Belin:2016yll}. 
 Intuitively, such a large spectrum gap is
 believed to be produced due to strong interactions in the holographic CFT. 
 This large spectrum gap is necessary to explain the black hole entropy expected from the AdS/CFT via a modular transformation. For example, if we assume a pure gravity theory on AdS$_3$, the spectrum gap of the conformal dimension $\Delta=h+\bar{h}$ is expected to be $\Delta_{gap}=\frac{c}{12}$. 
 Moreover, a basic property of quantum chaos so called eigenstate thermalization hypothesis (ETH) \cite{ETH}  has been derived from further studies of conformal bootstrap relations \cite{Brehm:2018ipf,Romero-Bermudez:2018dim,Hikida:2018khg}
 (see also \cite{Kraus:2016nwo,Cardy:2017qhl} for earlier related works).

 In this paper, we study an analogous spectrum property 
 in holographic BCFTs and 
 its physical implications. A boundary conformal field theory (BCFT) is defined as a CFT on a manifold  with boundaries where a part of conformal symmetry is preserved \cite{Cardy:1989ir,Cardy:2004hm}. The gravity duals of BCFTs 
 (AdS/BCFT) can be
 obtained by inserting a class of end of the world branes in AdS backgrounds,
 which satisfy Neumann boundary conditions dual to the boundary conformal invariance  \cite{Karch:2000gx,Takayanagi:2011zk,AdSBCFT,NTU}. 
 End of the world branes play a crucial role in recent progress in   understanding quantum aspects of black holes.
 Indeed, a  class  of microstates  of a single sided  AdS black hole with  the fixed mass $M$   can be constructed by the insertions of  the end of the world branes
to  an  eternal  two sided  black hole with the same mass $M$, \cite{Hartman:2013qma,Kourkoulou:2017zaj, Cooper:2018cmb}. Such a dictionary between black hole microstates and regularized boundary states  plays key role in
deriving the Page curve from the bulk perspective, therefore   a resolution of black hole information paradox in the light of Island formula 
 \cite{Penington:2019npb,Almheiri:2019psf,Almheiri:2019hni,Penington:2019kki,Almheiri:2019qdq,Rozali:2019day}. We also refer to \cite{Chen:2020uac,Balasubramanian:2020hfs} for studies  utilizing such branes to compute the  entropy of Hawking radiation of higher dimensional black holes. 
 End of the world branes also play a crucial role to find holographic duals of moving mirrors \cite{Akal:2020twv,Kawabata:2021hac}, which mimic black hole evaporation processes.
 
 Motivated by this, we will investigate a chaotic property of holographic
 boundary states (or so called Cardy states \cite{Cardy:1989ir}), namely 
 the inner products of two (regularized) boundary states. We will see that the off diagonal elements of them are exponentially suppressed in holographic BCFTs:
\ba
\la \psi_a|\psi_b\lb=\delta_{ab}+O(e^{-S_{BS}/2}),
\ea
where $|\psi_a\lb$ denotes the regularized boundary states, labeled by $a$.
 The quantity $S_{BS}$ estimates the entropy of microstates spanned by the boundary states. This behavior is directly related to the large spectrum gap in the open string between the two different boundaries $a$ and $b$.
 
This paper is organized as follows. In section two, we give a very brief review of
BCFT and boundary states and examine the inner products of boundary states. We will also present examples of inner products in a few solvable CFTs. In section three, starting with a short review of AdS/BCFT, we will study the relation between the inner products and open string spectra 
in two dimensional holographic CFTs, using a gravity dual. In section four, we extend the holographic analysis in section three to higher dimensional holographic CFTs. In section five, we will discuss our results, in particular by comparing them with the recent interpretation of a gravitational background as an ensemble of microstates. In appendix A, we give an argument which shows the absence of traversable wormholes in a class of Lorentzian AdS/BCFT setups.

\section{BCFT and Inner Products of Boundary States}

A boundary conformal field theory (BCFT) is a CFT on a manifold with boundaries with a suitable boundary condition described below \cite{Cardy:2004hm}. An Euclidean $d$ dimensional CFT preserves a conformal symmetry $SO(d+1,1)$. We choose a boundary condition which preserves a subgroup $SO(d,1)$ for a BCFT. In two dimensions $d=2$, full  conformal symmetry is enhanced to by a pair of  infinite dimensional Virasoro algebras. In the presence of a conformal boundary, the chiral half of them is preserved as the symmetry of the system. A boundary state \cite{Cardy:1989ir} is a useful description of a BCFT in terms of a quantum state. In this section,
we start with a brief review of boundary states and study their inner products. 
Although we focus on boundary states in two dimensional CFTs below, we can generalize their basic definition to higher dimensional CFTs in an obvious way.

\subsection{Boundary States}

A boundary state is a quantum state created by making a hole.
Consider a two dimensional CFT on a cylinder as depicted in the left picture in Fig.\ref{cylinderfig}.
We describe this cylinder by the coordinate 
$(\tau,x)$, imposing the periodicity $x\sim x+2\pi$. The Hamiltonian in the Euclidean time $\tau$ direction is denoted by $H_c$ (closed string Hamiltonian) and is written as
\ba
H_c=L_0+\bar{L}_0-\frac{c}{12},
\ea
in terms of the Virasoro generators and the central charge $c$.

Adding a boundary along $\tau=0$ is described by placing a boundary state (or Cardy state) $|B_a\lb$, where 
$a$ labels different boundary conditions. A Cardy state \cite{Cardy:1989ir} is a linear combination of Ishibashi states $|I_k\lb$ \cite{Ishibashi:1988kg}, where 
$k$ labels all primary states in the CFT as
\ba
|B_a\lb=\sum_k c^a_k |I_k\lb. \label{CST}
\ea
The Ishibashi state $|I_k\lb$ is a state constructed from a 
linear  combination of descendant states on top of the primary state labeled by $k$
and has maximal quantum entanglement between the left-moving and right-moving sectors.
Ishibashi states are orthogonal to each other. The amplitude  of the  Euclidean time evolution  by $\beta/2$   between two such states  is computed as
\ba
\la I_k |e^{-\frac{\beta}{2}H_c}|I_l\lb=\delta_{kl}~ \chi_k(e^{-\frac{\beta}{2}}).
\ea
$\chi_k$ is the character for the primary $k$.

\subsection{Open-Closed Duality and Overlaps}

\begin{figure}
  \centering
  \includegraphics[width=8cm]{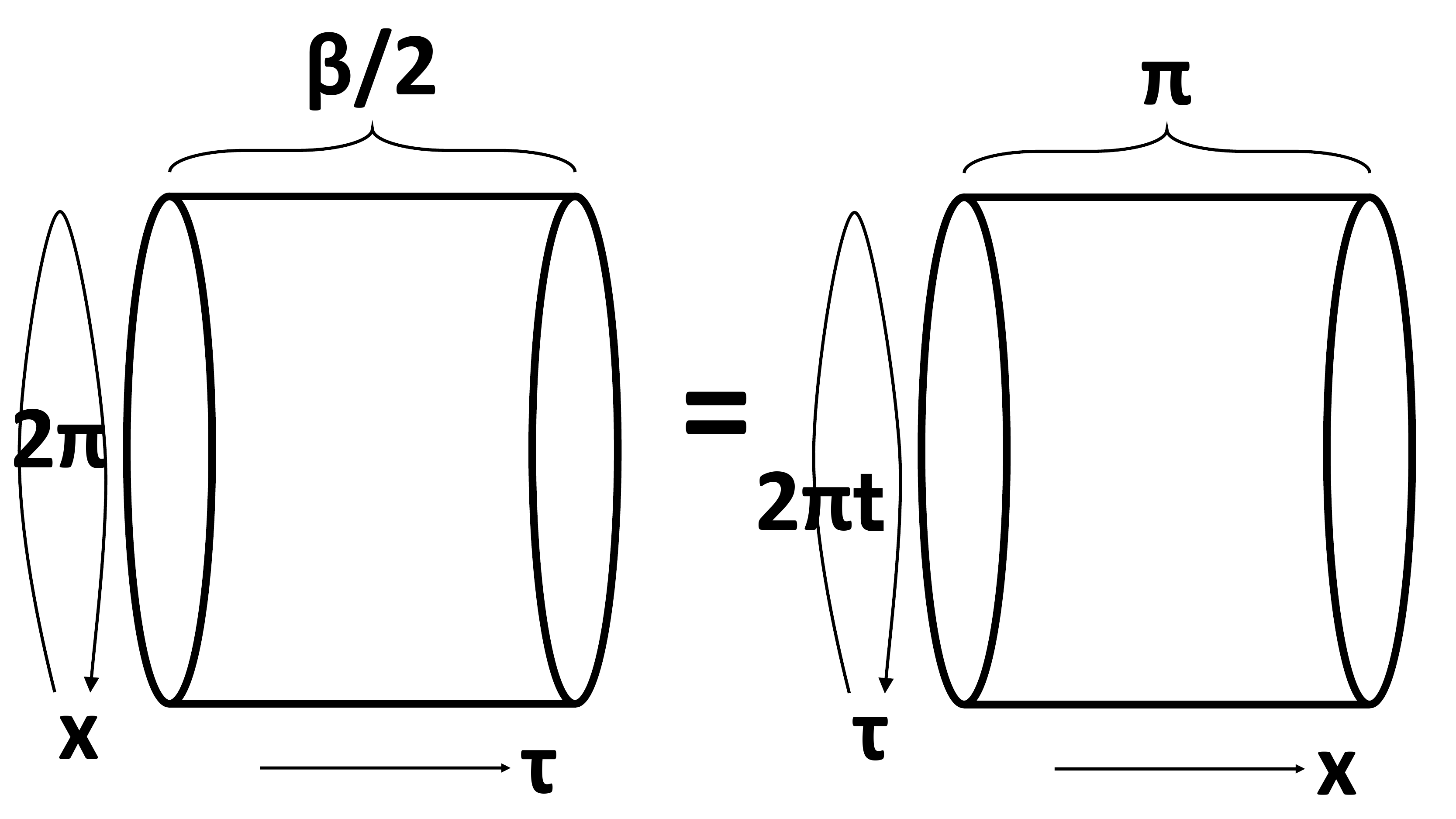}
  \caption{A sketch of cylinder amplitude from the viewpoint of closed string (left) and open string (right).}
\label{cylinderfig}
\end{figure}

On the other hand, the Cardy states are not  orthogonal to each other but satisfy the open-closed duality relation as follow (refer to Fig.\ref{cylinderfig})
\ba
\la B_a|e^{-\frac{\beta}{2}H_c}|B_b\lb=\sum_k N^{(k)}_{a,b} \mbox{Tr}_k \left[e^{-2\pi t H_{o}}\right],
\ea 
where 
\ba
\beta=\frac{2\pi}{t},
\ea
and $H_{o}=L_0-\frac{c}{24}$ is the open string Hamiltonian. In the right hand side, ${\rm Tr}_{k} [\cdots]$ denotes we take the trace with respect to the primary $k$ as well as its descendants. 
Moreover, $N^{(k)}_{a,b}$ counts the degeneracy of sectors which belong to the primary $k$
in the open strings between the boundaries $a$ and $b$.

Now if we take the limit  $\beta \to 0$ (or equally $t\to \infty$), we find
\ba
\la B_a|e^{-\frac{\beta}{2}H_c}|B_b\lb\simeq N^{(k_m)}_{a,b} 
e^{-2\pi t  \left(h^{(min)}_{a,b}-\frac{c}{24}\right)}, \label{opcd}
\ea
where $k_m$ is the lightest primary among those satisfy $ N^{(k_m)}_{a,b}\neq 0$, whose conformal 
dimension is denoted as $h^{(min)}_{a,b}$. 

Since we have $h^{(min)}_{a,a}=0$ when $a=b$,  
we can estimate the following ratio in the limit $\beta\to 0$ as
\ba
\frac{\la B_a|e^{-\frac{\beta}{2}H_c}|B_b\lb}{\s{\la B_a|e^{-\frac{\beta}{2}H_c}|B_a\lb\cdot \la B_b|e^{-\frac{\beta}{2}H_c}|B_b\lb}}\simeq N^{(k_m)}_{a,b} e^{-4\pi^2\frac{h^{(min)}_{a,b}}{\beta}}. \label{eq:overlapcft}
\ea
Note that here we used the fact $N_{a,a}^{(0)}=1$.

\subsection{Thermal Pure States}
One of the reasons we are interested in boundary states  is that, although they are pure states, in many respects they are indistinguishable from thermal mixed states under a suitable coarse-graining. This is in accord with the spirit of eigenstate thermalization hypothesis (ETH) \cite{ETH}, which claims  many properties of typical pure states coincide with those of  corresponding thermal mixed states.  Thus, one anticipates  that boundary states in CFT are indeed such typical states, therefore realize the ideas of ETH.  To make  this point explicit,
now we consider a pure state given by a regularized boundary state 
\ba
|\psi_a\lb={\cal N}_a \cdot  e^{-\frac{\beta}{4}H_c}|B_a\lb,
\label{micro}
\ea
where we assume $\beta$ is infinitesimally small and
we choose the normalization ${\cal N}_a$ such that $\la\psi_a|\psi_a\lb=1$.

This state is often employed as that just after a global quantum quench \cite{Calabrese:2005in}, which is an analytically  tractable model of thermalization in an isolated quantum system.
Thus, a late time limit of the time evolution of $|\psi_a\lb$ can be regarded as a thermal pure state. Indeed, its expectation value of the energy is computed as
\ba
E_{th}=\frac{\la\psi_a|H_c|\psi_a\lb}{\la\psi_a|\psi_a\lb}=\frac{\pi^2 c}{3\beta^2},
\label{eq:thenergy}
\ea
which agrees with the energy expectation value of the thermal ensemble with the temperature $1/\beta$. This indeed suggests the regularized boundary states are typical states \eqref{micro} with the temperature $\beta$.

However,   we can immediately see that not all typical states  with the fixed energy \eqref{eq:thenergy} belong  to the class of states\eqref{micro},  by counting the number of such regularized boundary states.The total number of the typical states  can be read off from the thermodynamic entropy, 
\ba
\# \mbox{Typical  states}\sim e^{S_{th}}, \quad
S_{th}=\frac{2\pi^2 c}{3\beta}.  \label{sth}
\ea


Now, let us also estimate the number of such  thermal pure states of the form (\ref{micro}), by counting the number of Cardy states.
Since the label $k$ of Ishibashi states coincides with that of primary states 
with the constraint 
that left and right are the same primary.
Therefore, the number of different Cardy states at an effective temperature $1/\beta$ is estimated as 
\ba
\# \mbox{Cardy states}\sim e^{S_{BS}}=e^{\frac{\pi^2c}{3\beta}},  \label{estimatem}
\ea 
where $S_{BS}=\frac{\pi^2c}{3\beta}$ denotes the  entropy naively associated with the  Cardy states. Thus we see  that the number of the regularized boundary states are too small to account  all typical states.

It is also useful to note that
in the AdS/CFT, this pure state (\ref{micro}) is dual to a microstate of a single sided BTZ black hole 
at inverse temperature $\beta$ \cite{Hartman:2013qma}, 
or equally a three dimensional 
AdS spacetime with an end of the world-brane where boundary conformal invariance is preserved \cite{Karch:2000gx,Takayanagi:2011zk}. This is indeed a half of the eternal BTZ black hole geometry which is dual to a thermal CFT.

\subsection{Inner Products}

Now we would like to evaluate the inner products $\la \psi_a|\psi_b\lb$ of the pure states
constructed from boundary states.
In the high temperature limit $\beta\to 0$, 
we can estimate their inner products by using (\ref{eq:overlapcft}) as follows:
\ba
\la\psi_a|\psi_b\lb\simeq \delta_{ab}+  N^{(k_m)}_{a,b} \cdot  e^{-\frac{4\pi^2}{\beta}  
h^{(min)}_{a,b}}. \label{bsa}
\ea
In this way, the larger gap in the open string channel leads to a larger exponential suppression of off diagonal elements of inner products.

As we will explain in the next section, for an ideal holographic CFT which is dual to a pure gravity on AdS$_3$, we expect for $a\neq b$
\ba
h^{(min)}_{a,b}=\frac{c}{24}.  \label{boundo}
\ea 
This bound may be regarded as a chiral version of the well-known maximal gap $\Delta=\frac{c}{12}$
for a CFT dual of a pure gravity on AdS$_3$ \cite{Hellerman:2009bu,Hartman:2014oaa}. 

If we introduce a overall random phase for the state 
as $|\psi_a\lb\to e^{i\theta_a}|\psi_a\lb$, we find that the inner product takes the following behavior 
\ba
\la\psi_a|\psi_b\lb\simeq\delta_{ab}+ R_{ab}\cdot e^{-\frac{S_{BS}}{2}} \label{beth},
\ea
where $S_{BS}$ is the same as (\ref{estimatem}). 
Also $R_{ab}$ is the  random fluctuations such that $\la R^*_{ab}R_{ab}\lb=1$.
This ETH like property implies that such states 
are random states among $e^{S_{BS}}$ states and thus 
nicely agrees with the estimation (\ref{estimatem}). 

This  should be compared to  the statistical fluctuations of overlaps between two typical states 
$| \phi_{i} \rangle$ \cite{Penington:2019kki}, 
\be
\langle \phi_{i} | \phi_{j} \rangle = \delta_{ij} + e^{-S_{{th}}/2}  R_{ij}, \label{eq:fueth}
\ee
with  the random variable $R_{ij}$ again satisfying  $\langle R^*_{ij} R_{ij} \rangle=1$. 
One can easily see the difference between the two, namely $S_{BS}$  in the overlap between two regularized boundary states (\ref{beth}) is replaced to the thermodynamic entropy $S_{{\rm th}} =2 S_{BS}$ in \eqref{eq:fueth}. This is again due to the fact that although such regularized boundary states are the frequently used model of thermal typical states,  the number of such states are quite small compared to the total number of typical states.

\subsection{Examples: Free Scalar and Liouville CFT}

There are several boundary states in two dimensional conformal field theories,   whose explicit  forms are known. In this section, we first compute the overlaps of such boundary states in free boson theory and  we  compare the results with the general  formula \eqref{eq:overlapcft} presented above. For completeness, we also list several known results in Liouville theory,  with caution that the general formula \eqref{eq:overlapcft} cannot be applied directly to this theory due to its non-unitarity and continuous spectrum.   Notice that these integrable boundary states have different properties than those of holographic boundary states.

\subsubsection{Boundary states in free boson theory}
Let us first consider the  boundary states in free boson theory, whose action is given by 
\be
I= \f{1}{2\pi} \int dz^{2} \partial_{z} X \partial_{\bar{z}} X.
\ee

There are two types of such states.   One is the Dirichlet state $| B_{D,x} \rangle $,  which is labeled by a real parameter $x$ of the localized point, 
and the other is  the  Neumann state $| B_{N} \rangle $. 

The overlap between two Dirichlet  boundary states $| B_{D,x_{1}} \rangle $ 
$| B_{D,x_{2}} \rangle $ is estimated as
\be
\f{\la B_{D,x_{2}} | e^{-\pi s H_{c}} |B_{D, x_{1}} \rangle }{\s{\la B_{D,x_{1}} | e^{-\pi s H_{c}} |B_{D, x_{1}} \rangle  \cdot \la B_{D,x_{2}} | e^{-\pi s H_{c}} |B_{D, x_{2}} \rangle }} = e^{-\f{(x_{1} -x_{2})^{2}}{4\pi s}}=e^{-\frac{(x_1-x_2)^2}{2\beta}},   \label{ddopen}
\ee
where we defined $s=\f{\beta}{2\pi}$. The above exponential suppression 
agrees with the lowest open string energy $h^{(min)}_{DD}=\frac{(x_1-x_2)^2}{8\pi^2}$ between 
the two boundaries, in accord with the general formula \eqref{eq:overlapcft}. 

The overlap between the Neumann state and a Dirichlet state is, 
\be
\f{\la B_{N} | e^{-\pi s H_{c}} |B_{D, x_{1}} \rangle }{\s{\la B_{N} | e^{-\pi s H_{c}} |B_{N} \rangle \cdot \la B_{D,x_{1}} | e^{-\pi s H_{c}} |B_{D, x_{1}} \rangle }} = 
\s{\f{4\pi \s{s} \eta^{3}(is) }{\theta_{2} (is)}}   \label{dnopen}
\ee
where the right hand  side is given by 
where $\eta(\tau)$ is the eta function, 
\be
\eta (\tau) =q^{\f{1}{24}} \prod^{\infty}_{n=1} (1-q^{n} ), \quad q=e^{2\pi i\tau} \quad 
\theta_{2} (\tau) =\prod_{n=1}^{\infty} (1-q^{n})(1+q^{n})^{2}.
\ee
In the $\beta\to 0$ limit, this ratio (\ref{dnopen}) behaves as 
$e^{-\frac{\pi^2}{4\beta}}$. This behavior agrees with the general formula \eqref{eq:overlapcft}, where the lowest open string energy is given by 
$h^{(min)}_{DN}=\frac{1}{16}$ between the Dirichlet and Neumann boundary.

\subsubsection{Boundary states in Liouville theory}  
Let us discuss a less trivial example, namely boundary states in Liouville theory. On a curved background its action reads, 
\be 
I=\f{1}{4\pi} \int dx^{2} \s{g}\; \left[  g^{ab} \p_{a} \phi \p_{b} \phi + Q R \phi + 4\pi  \mu e^{ 2b\phi}\right] ,
\ee
with $Q= b+ \f{1}{b}$, and $R$ is the Ricci scalar. The central charge of this CFT is $c= 1+6Q^{2}$. In particular, we are interested in the large $c$ limit $b\to 0$, where
$c\simeq \frac{6}{b^2}$.

This theory has  two types of boundary states, namely FZZT states \cite{FZZb,Tb} and ZZ \cite{ZZb} states. Since the overlaps between two FZZT states are divergent even in the presence of the UV regulator,  here we only discuss ZZ boundary states. ZZ states are characterized by two positive integers $(m,n)$, so we denote them  $|B_{(m,n)} \rangle$. The amplitude between two such states was computed in \cite{ZZb}:
\begin{align}
  \la B_{(m,n)}| e^{\pi i \tau H_{(m,n) (m',n')}} | B_{(m',n')} \rangle
   &=\sum_{k=0}^{{\rm min}(m,m')-1} \sum_{l=0}^{{\rm min}(n,n')-1} \chi_{m+m'-2k-1, n+n'-2l-1}(\tilde{q}),
\end{align}
where 
\be
\tau = \f{i \beta}{2\pi}, \quad \tilde{\tau}= -\f{1}{\tau}, \quad \tilde{q}= e^{2\pi i \tilde{\tau}}=e^{-\f{4\pi^{2}}{\beta}},
\ee
with the caution that we are using  an unusual convention for $\tau$, in order to match the notation with \eqref{eq:overlapcft}. Also, we introduced the degenerate character,
\be
\chi_{m,n}(q) =\f{q^{- \f{1}{4}\left(m/b+nb\right)^{2}}-q^{ -\f{1}{4}\left(m/b-nb\right)^{2}} }{\eta(\tau)}
\ee

For concreteness,  let us focus on three simplest states in this class, ie  $|B_{(1,1)}\rangle$,  $|B_{(1,2)}\rangle$, and $|B_{(2,1)}\rangle$.

The overlap between  $| B_{(1,1)} \rangle$ and $| B_{(1,2)} \rangle$  is 
\be
\f{\langle B_{(1,1)}| e^{-\pi s H_{c}} | B_{(1,2)} \rangle}{\s{\langle B_{(1,1)}| e^{-\pi s H_{c}} | B_{(1,1)} \rangle \langle B_{(1,2)}| e^{-\pi s H_{c}} | B_{(1,2)} \rangle}} =\f{\chi_{1,2}}{\s{\chi_{1,1 }(\chi_{1,1} +\chi_{1,3})}}.
\ee
When we take $\beta \rightarrow 0$ limit and the large $c$ limit $b\to 0$, we find 
 \be
 \la\psi_{(1,1)}|\psi_{(1,2)}\lb\simeq 1.
 \ee
This means that  $| B_{(1,1)} \rangle$ and $| B_{(1,2)} \rangle$  are indistinguishable in this limit. 
 
On the other hand the overlap between  $| B_{(1,1)} \rangle$ and $| B_{(2,1)} \rangle$  is 
\be
\f{\langle B_{(1,1)}| e^{-\pi s H_{c}} | B_{(2,1)} \rangle}{\s{\langle B_{(1,1)}| e^{-\pi s H_{c}} | B_{(1,1)} \rangle \langle B_{(2,1)}| e^{-\pi s H_{c}} | B_{(2,1)} \rangle}} =\f{\chi_{2,1}}{\s{\chi_{1,1 }(\chi_{1,1} +\chi_{3,1})}},  \quad s=\f{\beta}{2\pi}.
\ee
If we take the semi classical limit $b \rightarrow 0$ and the high temperature limit $\beta \rightarrow 0$  at the same time,  we get, 
\be
\la \psi_{(1,1)}|\psi_{(2,1)} \rangle \simeq e^{-\f{\pi^{2}c}{6\beta}},
\ee
which interestingly coincides with the expectation in holographic CFTs. 

More generally, in the large $c$ limit $b\to 0$, we obtain
\be
\la \psi_{(n_1,m_1)}|\psi_{(n_2,m_2)} \rangle \simeq e^{-\f{\pi^{2}c}{6\beta}(n_1-n_2)^2}.
\ee

Note that here we cannot apply directly the general formula \eqref{eq:overlapcft} to Liouville theory because Liouville theory is non-unitarity and has continuous spectrum, as opposed to the usual CFTs which we assumed in the derivation of \eqref{eq:overlapcft}. Here we nevertheless mention these states  because they are  examples of  boundary states  whose explicit form are known.

\section{Holographic Analysis in AdS$_3/$BCFT$_2$}

Now we move on to our main target: holographic BCFTs \cite{Takayanagi:2011zk,AdSBCFT,Sully:2020pza}. In this section, we focus on
two dimensional BCFTs with classical gravity duals. As we explained in the previous 
section, the presence of a conformal boundary is described by a Cardy state, labeled by 
boundary conditions. 

\subsection{Lightning Review of AdS/BCFT}

We apply the AdS/BCFT duality \cite{Takayanagi:2011zk,AdSBCFT,NTU} to study 
a gravity dual of a holographic CFT on a two-dimensional cylinder. 
The action of the gravitational system is given by 
\be
I=-\f{1}{16\pi G_N} \int_{N} dx^{3} \s{g} (R-2 \Lambda) -\f{1}{8\pi G_{N} }\int_{Q} \s{h} (K-T).
\ee
In the above action, $Q$ is the world volume of the end of the world brane in the bulk, which is anchored to the boundary of the BCFT region
at the asymptotic  boundary of the bulk spacetime. Also, in the action $K$ is the trace of the extrinsic curvature of $Q$, and $T$ denotes the tension of the brane.  
When we consider a gravity dual of a cylinder, there are two candidates of classical gravity solutions depending on whether the end of the world brane is connected or disconnected, as depicted in Fig.\ref{AdSBCFTfig}.
We call these two a connected and disconnected solution, respectively. 

First we consider a connected solution based on a thermal AdS$_3$.
We write the thermal AdS$_3$ metric as follows
\ba
ds^2=R^2\left(\frac{d\tau^2}{z^2}+\frac{dz^2}{h(z)z^2}+\frac{h(z)}{z^2}dx^2\right),
\ea
where
\ba
h(z)=1-\left(\frac{z}{z_0}\right)^2.
\ea
To make the geometry smooth, the coordinate $x$ is compactified as $x\sim x+2\pi z_0$.
We also compactify the Euclidean time $\tau$ such that $\tau\simeq \tau+2\pi z_H$.  
A gravity dual of a CFT on a cylinder is given by the subregion \cite{Takayanagi:2011zk,AdSBCFT}
\ba
z_0\arctan\left(\frac{RTz}{z_0\s{h(z)-R^2T^2}}\right)\leq x(z)\leq
\pi z_0-z_0\arctan\left(\frac{RTz}{z_0\s{h(z)-R^2T^2}}\right).
\ea
Note that at the AdS boundary $z=0$, this leads to $0\leq x\leq \pi z_0$ and this interval 
is identified with the one where the BCFT is defined.
The surface $Q$ which is the boundary of the above region is a connected surface.
By evaluating the gravity action on this background, we can holographically compute the BCFT partition function as follows
\ba
Z_{con}=e^{\frac{\pi cz_H}{12z_0}}.
\ea
Note that the result is independent from the value of the tension $T$, which parameterizes different boundary conditions.

Next we consider the disconnected solution based on the BTZ black hole solution
\ba
ds^2=R^2\left(\frac{f(z)d\tau^2}{\tau^2}+\frac{dz^2}{f(z)z^2}+\frac{dx^2}{z^2}\right),
\ea
where 
\ba
f(z)=1-\frac{z^2}{z_H^2}.
\ea
Now, in this solution, the Euclidean time like direction  contractible, as opposed to thermal AdS.

A gravity dual of a CFT on a cylinder is given by the region \cite{Takayanagi:2011zk,AdSBCFT}
\ba
-z_H\cdot \mbox{arcsinh}\left(\frac{R T_a z}{z_H\s{1-R^2T_a^2}}\right)\leq x(z)\leq \pi z_0+z_H\cdot
 \mbox{arcsinh}\left(\frac{R T_b z}{z_H\s{1-R^2T_b^2}}\right),
\ea
where $T_{a,b}$ is the tensions of each brane. By evaluating the gravity 
action on this background and we can again evaluate  the BCFT partition function as follows
\ba
&& Z_{dis}=e^{-I_{dis}},  \no
&&  I_{dis}=-\frac{\pi cz_0}{6z_H}-S^{(a)}_{bdy}-S^{(b)}_{bdy},
\ea
where $S^{(i)}_{bdy},$ $i=a,b$ are the boundary entropies \cite{Affleck:1991tk}
\ba
S^{(i)}_{bdy}=\frac{c}{3}\mbox{arctanh}(RT_i).
\ea 

\subsection{Inner Products of Holographic Boundary States}

\begin{figure}
  \centering
  \includegraphics[width=10cm]{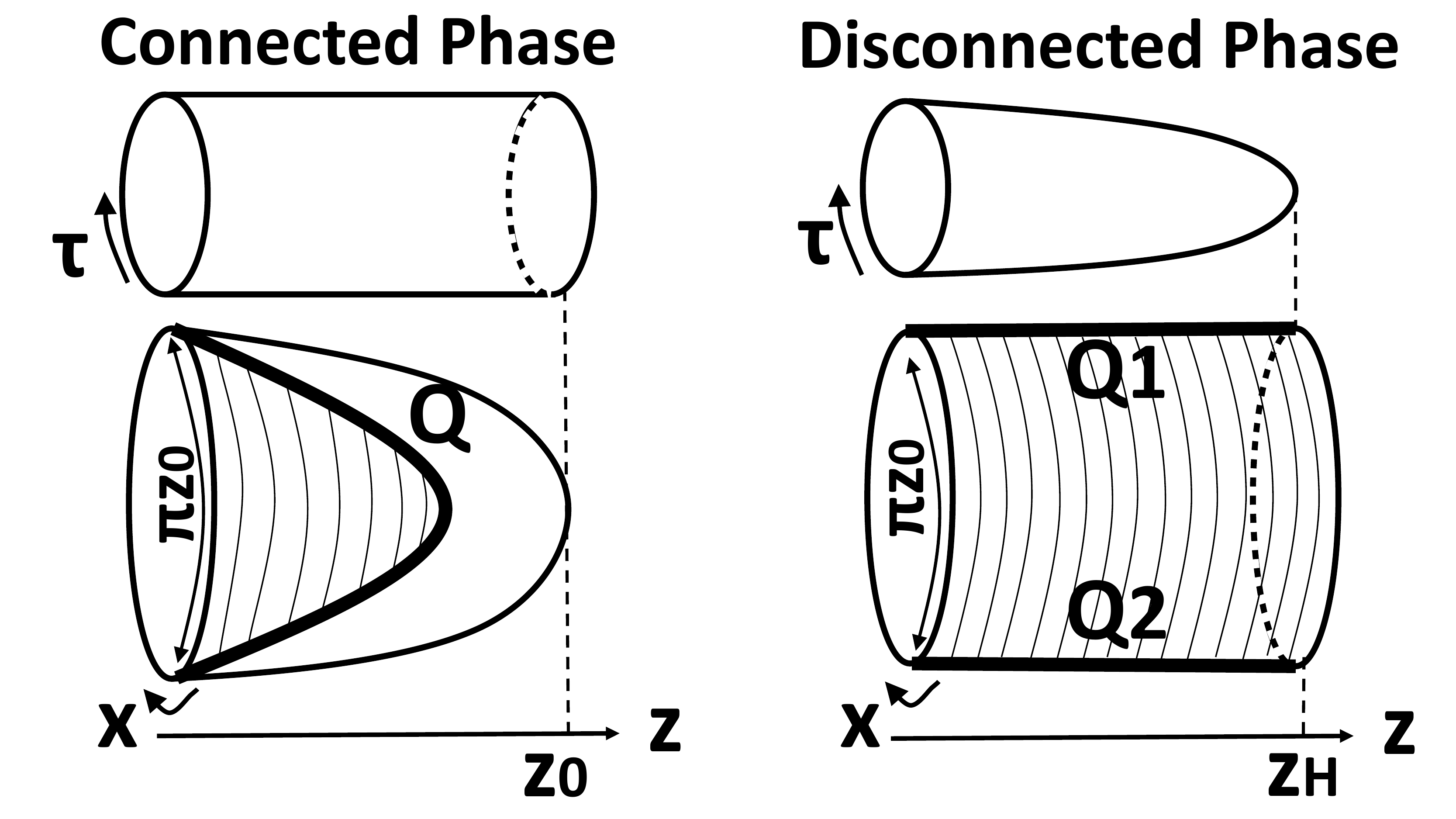}
  \caption{A sketch of gravity duals of a CFT on a cylinder in the connected phase (left) and disconnected phase (right) at $d=2$.}
\label{AdSBCFTfig}
\end{figure}

When we consider the overlap $\la B_a|e^{-\frac{\beta}{2}H_c}|B_a\lb$ for an identical boundary condition $a$, then both 
the connected and disconnected solutions are allowed, which are depicted as the left and right picture 
in Fig.\ref{AdSBCFTfig}. The connected solution is favored in the limit $\beta\to 0$. 
By choosing $\pi z_0=\frac{\beta}{2}$ and $z_H=1$ in order to adjust the normalization, we find 
\ba
\la B_a|e^{-\frac{\beta}{2}H_c}|B_a\lb\simeq e^{\frac{\pi^2 c}{6\beta}}.
\ea

When we consider $\la \psi_a|\psi_b\lb$ for two different boundary conditions $a$ and $b$, only the disconnected 
solution is allowed (i.e. the right picture in Fig.\ref{AdSBCFTfig}). Thus we can evaluate the overlap as follows
\ba
\la B_a|e^{-\frac{\beta}{2}H_c}|B_b\lb\simeq e^{\frac{c\beta}{12}+S^{(a)}_{bdy}+S^{(b)}_{bdy}}.  \ \ \  (a\neq b).
\ea

Thus in the limit $\beta\to 0$ we find that the inner products of the normalized pure state
$|\psi_a\lb$ (\ref{micro}) are given by as follows (when $a\neq b$)
\ba
\la \psi_a|\psi_b\lb
\simeq e^{-\frac{\pi^2 c}{6\beta}+S^{(a)}_{bdy}+S^{(b)}_{bdy}}.
\ea

This shows that the lowest dimensional state in the open string between $a$ and $b$ ($a\neq b$) is given by the previous expectation  (\ref{boundo}) i.e. $h^{(min)}_{a,b}=\frac{c}{24}$, via (\ref{bsa}), as promised.

\section{Higher Dimensional Generalizations}

The holographic approach based on the AdS/BCFT provides predictions also in higher dimensional BCFTs \cite{AdSBCFT}. Notice that we can define a boundary state $|B_a\lb$ as a state making a $d-1$ dimensional hole in $d$ dimensional space. 
The inner product $\la B_a|e^{-\frac{\beta}{2}H}|B_b\lb$ 
can be computed as a partition function on 
a $d$ dimensional open manifold $I_{\beta/2}\times T^{d-1}$, where $I_{\beta/2}$ is a length $\beta/2$ interval. As in the previous $d=2$ case, there are two different solutions, depending on whether the end of the world brane is connected or disconnected. 

First we consider a connected solution based on a AdS$_{d+1}$ soliton.
We write the AdS$_{d+1}$ soliton geometry as follows
\ba
ds^2=R^2\left(\frac{d\tau^2+\sum_{i=1}^{d-2}dy_i^2}{z^2}+\frac{dz^2}{h(z)z^2}+\frac{h(z)}{z^2}dx^2\right),
\ea
where
\ba
h(z)=1-\left(\frac{z}{z_0}\right)^d.
\ea
We compactify $x\sim x+\frac{4\pi}{d}z_0$ in order to have a smooth geometry. We also compactify the Euclidean time $\tau\sim \tau+\frac{4\pi}{d}z_H$ and other spacial coordinates $y_i\sim y_i+L_i$. The boundary CFT is defined in the bounded region  
\ba
 |x|\leq x(T).
\ea
The function $x(T)$ is introduced as
\ba
&& x(T):=\xi(T)\cdot z_0,
\ea
where $\xi(T)$ is defined when $T\leq 0$ as follows \cite{AdSBCFT}:
\ba
 \xi(T)\equiv \frac{\Gamma(1/d)\Gamma(1/2)}{\Gamma(1/2+1/d)}\frac{R|T|}{d(d-1)}\left(1-\frac{R^2T^2}{(d-1)^2}\right)^{1/d-1/2}F\left(1,\frac{1}{d},\frac{1}{2}+\frac{1}{d};1-\frac{R^2T^2}{(d-1)^2}\right).\nonumber
\ea
When $T>0$, it is defined by $\xi(T)=\frac{2 \pi}{d}-\xi(-T)$. Note that the tension takes values in the range $|T|< \frac{d-1}{R}$.
As we see in the previous section, when we set $d=2$ we have $\xi(T)=\frac{\pi}{2}$ for any $T$. 

We would like to point out that in higher dimensions the behavior of $\xi(T)$ is different from that in $d=2$. For $d>2$, $\xi(T)$ non-trivially depends on $T$. We find $\xi(T)$ is a monotnically decreasing function of $T$ such that 
$\xi\left(-\frac{d-1}{R}\right)=\infty$, $\xi(0)=\frac{\pi}{d}$ and $\xi(T_*)=0$. Here $T_*>0$ depends on the dimension. Since we have $\xi<0$ for $T>T_*$, which looks unphysical, this implies that there is an upper bound of the tension $T<T_*$ for $d>2$.

By evaluating the on-shell gravity action, the partition function is obtained as \cite{AdSBCFT} 
\ba
Z_{con}=\text{e}^{\frac{R^{d-1}(\prod_{i=1}^{d-2} L_i) z_H x(T)}{2d G_Nz_0^d}}.
 \label{pacon}
\ea

Next we consider the disconnected solution based on the AdS Schwartzshild black hole solution
\ba
ds^2=R^2\left(\frac{f(z)d\tau^2}{z^2}+\frac{dz^2}{f(z)z^2}+\frac{dx^2+\sum_{i=1}^{d-1}dy_i^2}{z^2}\right),
\ea
where 
\ba
f(z)=1-\left(\frac{z}{z_H}\right)^d.
\ea
We compactify $\tau\sim \tau+\frac{4\pi}{d}z_H$ in order to have a smooth geometry. We also compactify other spacial coordinates $y_i\sim y_i+L_i$, and the boundaries sit at $x=\pm x(T)$. For simplicity, we consider 
the case where the tension of the surface $Q$ is vanishing $T=0$. This is enough to extract the leading behavior in the limit $\beta\to 0$ as was true in $d=2$. Then the disconnected solution is given by the region
\ba
|x|\leq x(T=0)=\frac{\pi}{d}z_0.
\ea
By evaluating the gravity action on this background and we can estimate the partition function as follows
\ba
&& Z_{dis}=e^{-I_{dis}},  \label{padis}\\
&&  I_{dis}=-\frac{R^{d-1}(\prod_{i=1}^{d-2} L_i) x(T=0)}{dG_Nz_H^{d-1}},
\ea

Now let us evaluate the inner products of boundary states using the above holographic results.
For an identical boundary condition $a$ i.e. 
$\la B_a|e^{-\frac{\beta}{2}H_c}|B_a\lb$, both the connected and disconnected solutions are allowed as in AdS${}_3$. The connected solution is favored in the limit $\beta\to 0$. We set $2x(T)=2\xi(T)z_0=\frac{\beta}{2}$ and take the periodicity in the $\tau$ and $y_i$ direction to be $\frac{4\pi}{d}z_H=2\pi$ and $L_i=L$, respectively. Then we find from (\ref{pacon})
\ba
\la B_a|e^{-\frac{\beta}{2}H_c}|B_a\lb_{con}\simeq e^{(4\xi(T))^d\pi c_{\text{eff}}\frac{L^{d-2}}{\beta^{d-1}}}.
\ea
Here we defined 
\ba
c_{\text{eff}}:=\frac{R^{d-1}}{16\pi G_N}.
\ea
Note that when $d=2$, it is related to the central charge by $c_{\text{eff}}=\frac{c}{24\pi}$. 

When we consider an inner product for two different boundary conditions $a$ and $b$, only the disconnected 
solution is allowed (i.e. the right picture in Fig.\ref{AdSBCFTfig}). Thus we can evaluate the overlap for $T=0$ from (\ref{padis}) as follows
\ba
\la B_a|e^{-\frac{\beta}{2}H_c}|B_b\lb\simeq e^{\frac{8\pi}{ d^{d+1}}c_{\text{eff}}\beta L^{d-2}}.  \ \ \  (a\neq b).
\ea
Thus in the limit $\beta\to 0$ we find\footnote{Here we note that the $T$ dependence for the disconnected solution can be negligible when we focus on the leading behavior of this ratio in the limit $\beta\to 0$.}
\ba
\frac{\la B_a|e^{-\frac{\beta}{2}H_c}|B_b\lb}{\s{\la B_a|e^{-\frac{\beta}{2}H_c}|B_a\lb\cdot \la B_b|e^{-\frac{\beta}{2}H_c}|B_b\lb}}
\simeq e^{-\frac{\pi c_{\text{eff}}}{2}\frac{L^{d-2}}{\beta^{d-1}}\left((4\xi(T_a))^d+(4\xi(T_b))^d\right)},
\ea
where $T_a$ and $T_b$ are the tensions of surfaces $Q_a$ and $Q_b$ dual to the boundary states $|B_a\lb$ and 
$|B_b\lb$, respectively.

This predicts that the lowest energy among states in the open string between $a$ and $b$ ($a\neq b$) is given by \ba
E^{(min)}_{a,b}=\frac{\pi c_{\text{eff}}}{2}
\left[\left(\frac{2\xi(T_a)}{\pi}\right)^d+\left(\frac{2\xi(T_b)}{\pi}\right)^d\right]  V_{d-2},
\ea
where $V_{d-2}=\left(\frac{2\pi L}{\beta}\right)^{d-2}$ is the volume of $d-2$ dimensional torus in $y^i$ direction. We took the length of $x$ to be $\pi$ and the periodicity of $\tau$ to be $2\pi t=\frac{4\pi^2}{\beta}$ as in the right of Fig.\ref{cylinderfig}. Note also that the above energy gap decreasing as the tensions $T_a$ and $T_b$ gets larger
and does vanish when $T_a=T_b=T_*$.

\section{Discussions: JT gravity vs AdS$_3/$BCFT$_2$ ?}

In this paper, we studied quantum states in CFTs $|\psi_a\lb$, which are defined by regularized boundary states (Cardy states) for all possible conformal boundary conditions, labeled by $a$. This provides a class of  microstates for single sided black holes in AdS. Since the left-right symmetric constraint is imposed for boundary states, the number of microstates spanned by boundary states, denoted by $e^{S_{BS}}$,  is of the order of the square root of $e^{S_{th}}$ i.e. $S_{BS}\simeq \frac{S_{th}}{2}$. Here
$e^{S_{th}}$ denotes  the number of typical states $|\phi_{i}\lb$ in a CFT which are indistinguishable from the corresponding thermal mixed state.
$S_{th}$ is also equal to the entropy of  eternal black hole dual to the thermo field double of a holographic CFT.

\begin{figure}
  \centering
  \includegraphics[width=10cm]{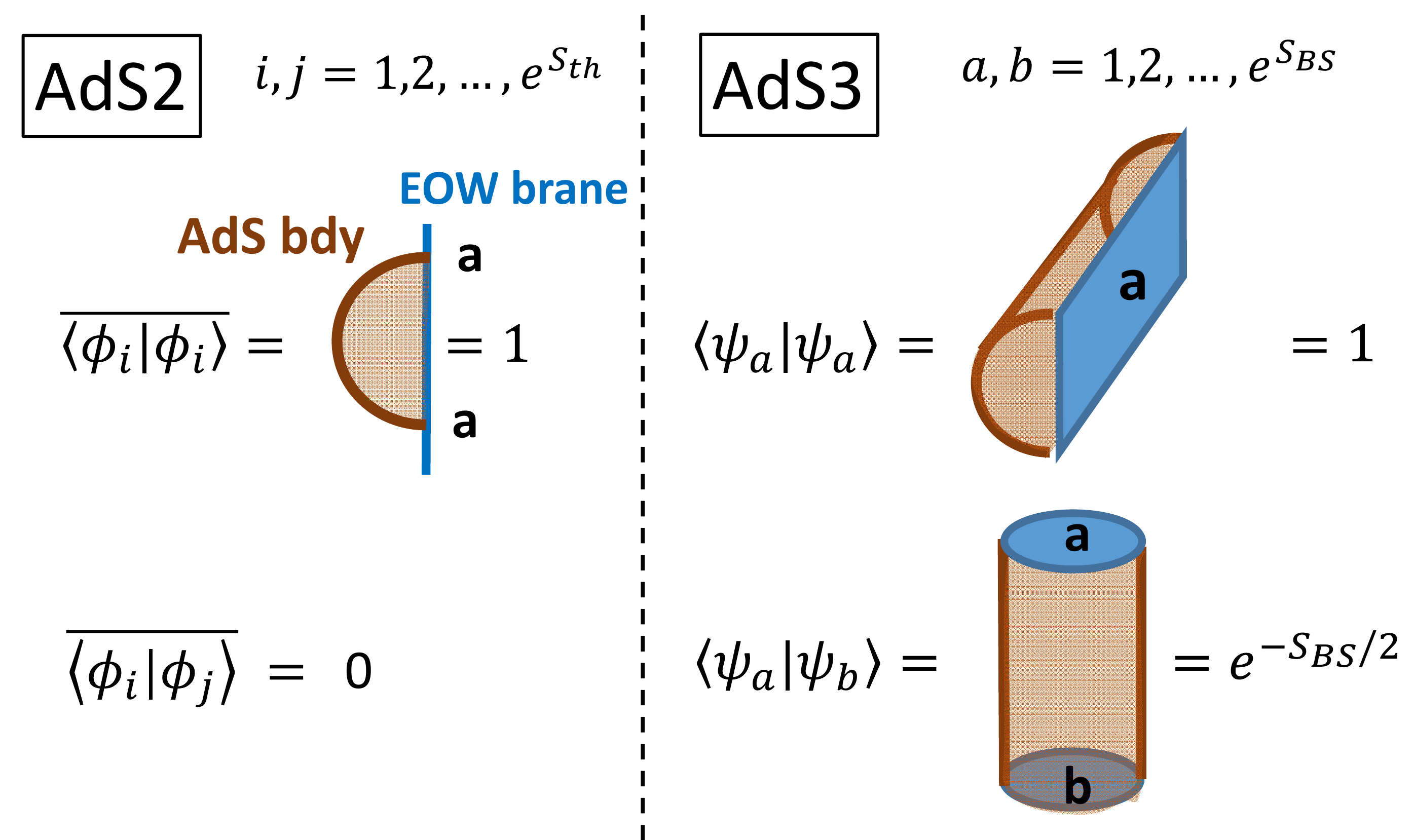}
  \caption{Gravity evaluations of the inner products of microstates  $\la \phi_i|\phi_j\lb$ in AdS$_2$ gravity
  (left) and those of regularized boundary states  $\la \psi_a|\psi_b\lb$  in  
   AdS$_3$ gravity (right).}
\label{wormholefig}
\end{figure}

By using the open-closed duality we showed that the inner products $\la \psi_a|\psi_b\lb$ of boundary states for $a\neq b$ are exponentially small such that its exponent is proportional to the lowest energy in the open string spectrum between $a$ and $b$ as in (\ref{bsa}). 
The holographic analysis of this inner products for a three dimensional AdS pure gravity shows that this lowest energy is $h^{(min)}_{ab}=\frac{c}{24}$ in its dual two dimensional holographic CFT. Indeed, this value corresponds the estimation of inner products: $\la \psi_a|\psi_b\lb\sim e^{-S_{BS}/2}$. A most crucial fact behind this estimation is that there are two solutions in AdS/BCFT depending on whether the end of the world brane (EOW brane) is connected or the EOW branes are disconnected, as depicted as the top and bottom picture in Fig.\ref{wormholefig}, respectively. This result means that the quantum states $|\psi_a\lb$ are 
random vectors in the full space spanned by the left-right symmetric states.  
Therefore we expect that this lowest energy $h^{(min)}_{ab}=\frac{c}{24}$ is maximum possible values for any two dimensional BCFTs. This implies that such CFTs correspond to maximally chaotic BCFTs and that holographic BCFTs are the most chaotic BCFTs. We also generalized the above analysis to higher dimensional BCFTs, by applying the AdS/BCFT. An important new aspect in higher dimensions is that the lowest energy in open string turns out to depend on the value of the tension of the EOW branes. Therefore it is an intriguing future problem to understand the distributions of values of tension for the EOM branes which are dual to conformal boundary states.

\begin{figure}
  \centering
  \includegraphics[width=10cm]{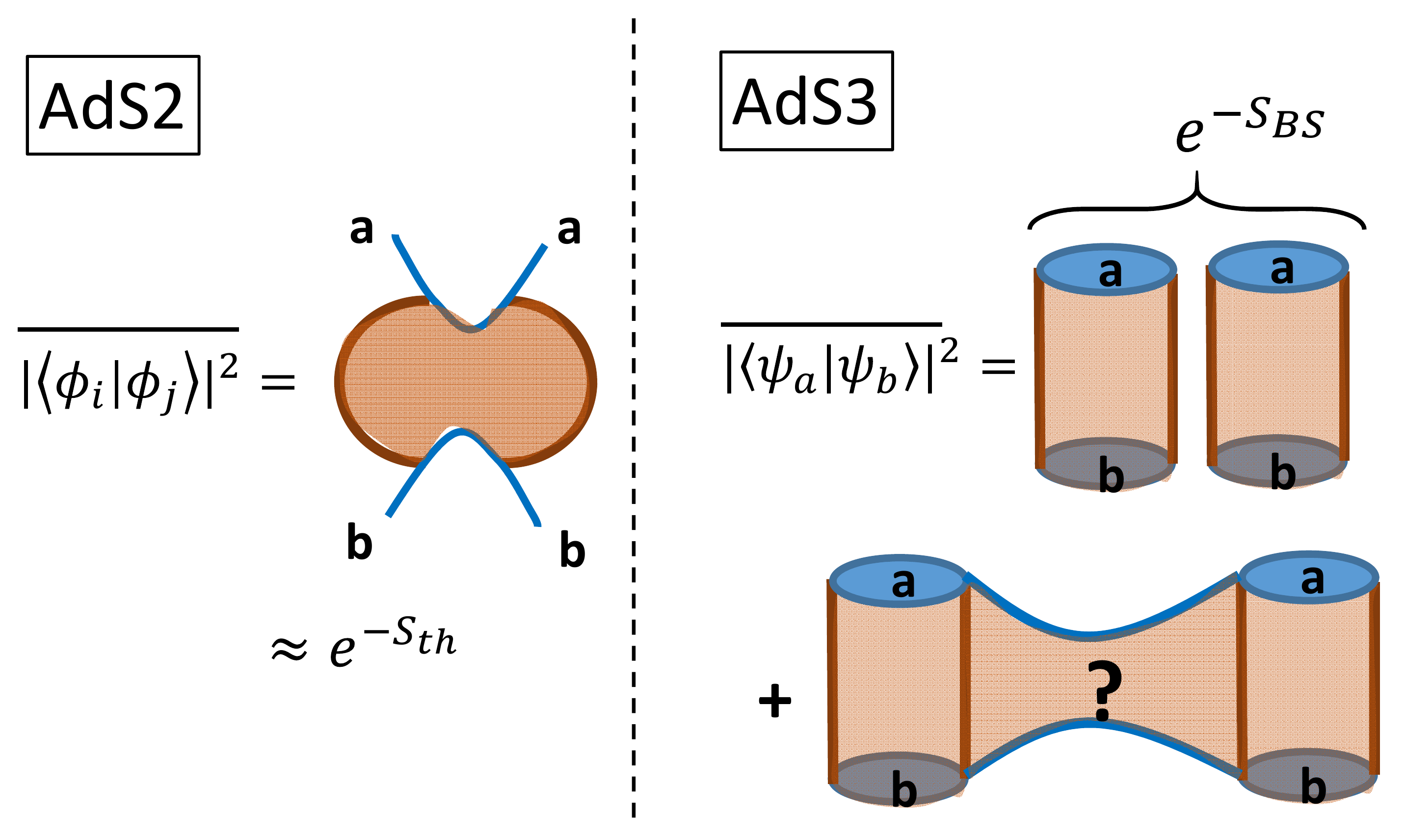}
  \caption{Gravity evaluations of 
 the averages of squares of inner products in AdS$_2$ gravity (left) and AdS$_3$ gravity (right).}
\label{wormholeefig}
\end{figure}

Now we would like to compare our results with the analysis of microstates in two dimensional AdS gravity (JT gravity) performed in \cite{Penington:2019kki}
to explain the physics behind the island formula in the black hole information loss problem. First of all, when we evaluate the inner products of microstates $\la \phi_i|\phi_j\lb$, they vanish except $a=b$ in the AdS$_2$ gravity as depicted in the left of Fig.\ref{wormholefig}. In \cite{Penington:2019kki}, this is interpreted that the gravity calculation corresponds to taking a random ensemble average $\ov{\la \phi_i|\phi_j\lb}$ with respect to all possible microstates $i,j$. Even though the gravity analysis cannot directly compute the off diagonal element of $\la \psi_i|\psi_j\lb$, its square average  
$\ov{|\la \phi_i|\phi_j\lb|^2}$ is argued to be computable in the two dimensional gravity, 
by taking wormhole solutions (called replica wormholes) into account, as depicted in the left of 
Fig.\ref{wormholeefig}. This leads to the evaluation
$\ov{|\la \phi_a|\phi_b\lb|^2}\sim e^{-S_{th}}$ and this implies the behavior $\la \phi_a|\phi_b\lb\simeq\delta_{ij}+R_{ij}e^{-\frac{S_{th}}{2}}$   \cite{Penington:2019kki}, where 
$R_{ij}$ is a random matrix. 

On the other hand, in the three dimensional AdS gravity we studied in this paper, the situation is a bit different.  Since we can have the solution with two disconnected EOW branes in AdS$_3$ as depicted in the lower right picture of  Fig.\ref{wormholefig}, the gravity prediction for the inner product $\la \psi_a|\psi_b\lb$  does not vanish even if $a \neq b$, as opposed to the AdS$_2$ case of  \cite{Penington:2019kki}.
Therefore we obtain the off diagonal value of 
$\la \psi_a|\psi_b\lb$ directly from a gravity calculation. 
This leads to the behavior 
$\la \psi_a|\psi_b\lb\simeq\delta_{ab}+O(e^{-\frac{S_{BS}}{2}})$, which leads to the lowest energy gap 
$h^{(min)}_{ab}=\frac{c}{24}$ of open strings.
These highlight the main difference between the microstate analysis in two dimensional JT gravity and that in three dimensional AdS gravity.

It may be plausible to think the ensemble average $\ov{|\la \psi_a|\psi_b\lb|^2}$ can be computed by the gravity partition function for a geometry with four boundaries i.e. two $a$s and two $b$s, which depicted in the right of Fig.\ref{wormholeefig}. The first term is the direct product of the cylinder solutions. We expect that the random averages are responsible for the second term which corresponds to wormhole geometries which connect two cylinders. However, when we consider the construction of Euclidean AdS wormholes which connect two boundaries given in \cite{Maldacena:2004rf}, each boundary has to be a surface with genus higher than one. In particular, we cannot connect two tori by an on shell AdS wormhole . Similarly, we might expect that it is not possible to connect two cylinders by an AdS wormhole with appropriate EOW branes.\footnote{In the appendix A, we gave an argument that we cannot construct a travesable wormhole in the Lorentzian AdS/BCFT, by picking up a class of examples.} This may suggest that in AdS$_3/$BCFT$_2$, 
we do not need to regard gravity path-integrals as averaged quantities of holographic CFTs. 
It would be an intriguing future problem to explore more on these aspects to understand precisely how gravity can describe CFT microstates.

\section*{Acknowledgements}

We are grateful to Raphael Bousso, Norihiro Iizuka, Akihiro Ishibashi, and Yoshifumi Nakata for useful discussions.
We would like to thank YITP online workshop "Recent progress in theoretical physics based on quantum information theory" (YITP-W-20-15) hosted by Yukawa Institute for Theoretical Physics, Kyoto U., where this work was completed. 
TT is supported by the Simons Foundation through the ``It from Qubit'' collaboration,  Inamori Research Institute for Science and World Premier International Research Center Initiative (WPI Initiative)
from the Japan Ministry of Education, Culture, Sports, Science and Technology (MEXT).
TT is supported by JSPS Grant-in-Aid for Scientific Research (A) No.~16H02182 and
by JSPS Grant-in-Aid for Challenging Research (Exploratory) 18K18766. MM is supported in part by the Berkeley Center for Theoretical Physics; by the Department of Energy, Office of Science, Office of High Energy Physics under QuantISED Award desc0019380 and under contract DE-AC02-05CH11231; and by the National Science Foundation under grant PHY1820912. TU was supported by JSPS Grant-in-Aid for Young Scientists  19K14716.

\appendix
\section{Averaged Null Energy Condition and Absence of Traversal Wormholes in AdS/BCFT}

Here we will show that a construction of traversal wormhole with a locally Poincare AdS metric is not possible in AdS/BCFT when we impose the averaged energy condition (ANEC). An analogous statement of the absence of traversable wormholes in the standard AdS/CFT without end of the world branes was proven in \cite{Gao:2000ga,Galloway:1999br}. The ANEC for quantum field theories on flat spaces was derived in \cite{Kelly:2014mra} using the AdS/CFT and in \cite{Faulkner:2016mzt,Hartman:2016lgu} using field theoretic arguments. Recently, a necessary modification of ANEC for spaces with positive curvatures was proposed in \cite{Iizuka:2019ezn} via holography.

In the Poincare AdS${}_{d+1}$
\ba
ds^2=R^2\left(\frac{-dt^2+dz^2+dx^2+\sum_{i=1}^{d-2}dy_i^2}{z^2}\right),  \label{pohig}
\ea
we consider two BCFTs on regions confined by $x\leq -a$ or $x\geq a$, and ask whether we can have a static EOW brane that ends at these two boundaries. We write $x$ coordinates of the brane at $x\leq 0$ by $x(z)$, in particular we have $x(0)=-a$ and $x'(z^*)=\infty$ where $z=z^*$ is the turning point. The null vector which generates null geodesic on the surface $Q$
\ba
\left(N^t,~N^z,~N^x\right)=Rz^2\left(-1,~\frac{1}{\sqrt{1+x'(z)^2}},~\frac{x'(z)}{\sqrt{1+x'(z)^2}}\right).
\ea
We fixed the normalization of null vector such that we have $\frac{dX^{\mu}}{d\lambda}=N^\mu$ for an affine parameter $\lambda$, namely it satisfies $N^\mu \nabla_\mu N^\nu=0$.

The averaged null energy can be decomposed into two parts: the one from $x=-a$ to $x=x(z_*)$ and the other one $x=x(z_*)$ to $x=a$. The averaged null energy condition tells us that either of them is non-negative, which we can take to be the first one without losing any generality. Therefore we require
\ba
\int~d\lambda \left(K_{\mu\nu}-Kh_{\mu\nu}\right)N^{\mu}N^{\nu}=-\int^{z^*}_{\epsilon}~dz\frac{zx''(z)}{(1+x'(z)^2)}\geq 0.
\ea
However we can show that this integral should be negative by performing a partial integration:
\ba
&& -\int^{z^*}_{\epsilon}~dz\frac{zx''(z)}{(1+x'(z)^2)}  \no
&&=\left[-z\cdot \arctan(x')\right]^{z_*}_{\ep}
+\int^{z_*}_0 dz \arctan x'=-\frac{\pi}{2}z_*+\int^{z_*}_0 dz \arctan x'<0,
\ea
where we employed that the fact we have $x'=\infty$ at the turning point $z=z_*$ and the bound
$|\arctan x'|< \frac{\pi}{2}$ for $\ep\leq z<z_*$. This clearly shows the wormhole solution in the geometry 
(\ref{pohig}) is not possible.



\begin{thebibliography}{99}


\bibitem{Ma}
  J.~M.~Maldacena,
  Adv.\ Theor.\ Math.\ Phys.\  {\bf 2} (1998) 231
  [Int.\ J.\ Theor.\ Phys.\  {\bf 38} (1999) 1113]
  [arXiv:hep-th/9711200];


\bibitem{Maldacena:2015waa}
J.~Maldacena, S.~H.~Shenker and D.~Stanford,
``A bound on chaos,''
JHEP \textbf{08} (2016), 106
[arXiv:1503.01409 [hep-th]].



\bibitem{Heemskerk:2009pn}
I.~Heemskerk, J.~Penedones, J.~Polchinski and J.~Sully,
``Holography from Conformal Field Theory,''
JHEP \textbf{10} (2009), 079
[arXiv:0907.0151 [hep-th]].


\bibitem{Hartman:2014oaa}
T.~Hartman, C.~A.~Keller and B.~Stoica,
``Universal Spectrum of 2d Conformal Field Theory in the Large c Limit,''
JHEP \textbf{09} (2014), 118
[arXiv:1405.5137 [hep-th]].



\bibitem{Belin:2016yll}
A.~Belin, J.~de Boer, J.~Kruthoff, B.~Michel, E.~Shaghoulian and M.~Shyani,
``Universality of sparse $d > 2$ conformal field theory at large $N$,''
JHEP \textbf{03} (2017), 067
[arXiv:1610.06186 [hep-th]].


\bibitem{ETH}
 M. Srednicki, 
 “The Approach to Thermal Equilibrium in Quantized Chaotic Systems,” J.
Phys. \textbf{A 32} (1999) 1163, cond-mat/9809360.

\bibitem{Brehm:2018ipf}
E.~M.~Brehm, D.~Das and S.~Datta,
``Probing thermality beyond the diagonal,''
Phys. Rev. D \textbf{98} (2018) no.12, 126015
[arXiv:1804.07924 [hep-th]].

\bibitem{Romero-Bermudez:2018dim}
A.~Romero-Berm\'udez, P.~Sabella-Garnier and K.~Schalm,
``A Cardy formula for off-diagonal three-point coefficients; or, how the geometry behind the horizon gets disentangled,''
JHEP \textbf{09} (2018), 005
[arXiv:1804.08899 [hep-th]].

\bibitem{Hikida:2018khg}
Y.~Hikida, Y.~Kusuki and T.~Takayanagi,
``Eigenstate thermalization hypothesis and modular invariance of two-dimensional conformal field theories,''
Phys. Rev. D \textbf{98} (2018) no.2, 026003
[arXiv:1804.09658 [hep-th]].

\bibitem{Kraus:2016nwo}
P.~Kraus and A.~Maloney,
``A cardy formula for three-point coefficients or how the black hole got its spots,''
JHEP \textbf{05} (2017), 160
[arXiv:1608.03284 [hep-th]].

\bibitem{Lashkari:2016vgj}
N.~Lashkari, A.~Dymarsky and H.~Liu,
``Eigenstate Thermalization Hypothesis in Conformal Field Theory,''
J. Stat. Mech. \textbf{1803} (2018) no.3, 033101
[arXiv:1610.00302 [hep-th]].

\bibitem{Cardy:2017qhl}
J.~Cardy, A.~Maloney and H.~Maxfield,
``A new handle on three-point coefficients: OPE asymptotics from genus two modular invariance,''
JHEP \textbf{10} (2017), 136
[arXiv:1705.05855 [hep-th]].


\bibitem{Cardy:1989ir}
J.~L.~Cardy,
``Boundary Conditions, Fusion Rules and the Verlinde Formula,''
Nucl. Phys. B \textbf{324} (1989), 581-596.



\bibitem{Cardy:2004hm}
J.~L.~Cardy,
``Boundary conformal field theory,''
[arXiv:hep-th/0411189 [hep-th]].



\bibitem{Karch:2000gx}
A.~Karch and L.~Randall,
``Open and closed string interpretation of SUSY CFT's on branes with boundaries,''
JHEP \textbf{06} (2001), 063
[arXiv:hep-th/0105132 [hep-th]].


\bibitem{Takayanagi:2011zk}
 T.~Takayanagi,
  ``Holographic Dual of BCFT,'' 
   Phys.\ Rev.\ Lett.\  {\bf 107} (2011) 101602.

\bibitem{AdSBCFT}
M.~Fujita, T.~Takayanagi and E.~Tonni,
``Aspects of AdS/BCFT,''
JHEP \textbf{11} (2011), 043
[arXiv:1108.5152 [hep-th]].

\bibitem{NTU}
M.~Nozaki, T.~Takayanagi and T.~Ugajin,
``Central Charges for BCFTs and Holography,''
JHEP \textbf{06} (2012), 066
[arXiv:1205.1573 [hep-th]].

\bibitem{Hartman:2013qma}
T.~Hartman and J.~Maldacena,
``Time Evolution of Entanglement Entropy from Black Hole Interiors,''
JHEP \textbf{05} (2013), 014
[arXiv:1303.1080 [hep-th]].

\bibitem{Kourkoulou:2017zaj}
I.~Kourkoulou and J.~Maldacena,
``Pure states in the SYK model and nearly-$AdS_2$ gravity,''
[arXiv:1707.02325 [hep-th]].

\bibitem{Cooper:2018cmb}
S.~Cooper, M.~Rozali, B.~Swingle, M.~Van Raamsdonk, C.~Waddell and D.~Wakeham,
``Black Hole Microstate Cosmology,''
JHEP \textbf{07}, 065 (2019)
doi:10.1007/JHEP07(2019)065
[arXiv:1810.10601 [hep-th]].



\bibitem{Penington:2019npb}
G.~Penington,
``Entanglement Wedge Reconstruction and the Information Paradox,''
JHEP \textbf{09} (2020), 002
[arXiv:1905.08255 [hep-th]].

\bibitem{Almheiri:2019psf}
A.~Almheiri, N.~Engelhardt, D.~Marolf and H.~Maxfield,
``The entropy of bulk quantum fields and the entanglement wedge of an evaporating black hole,''
JHEP \textbf{12} (2019), 063
[arXiv:1905.08762 [hep-th]].

\bibitem{Almheiri:2019hni}
A.~Almheiri, R.~Mahajan, J.~Maldacena and Y.~Zhao,
``The Page curve of Hawking radiation from semiclassical geometry,''
JHEP \textbf{03} (2020), 149
[arXiv:1908.10996 [hep-th]].

\bibitem{Penington:2019kki}
G.~Penington, S.~H.~Shenker, D.~Stanford and Z.~Yang,
``Replica wormholes and the black hole interior,''
[arXiv:1911.11977 [hep-th]].

\bibitem{Almheiri:2019qdq}
A.~Almheiri, T.~Hartman, J.~Maldacena, E.~Shaghoulian and A.~Tajdini,
``Replica Wormholes and the Entropy of Hawking Radiation,''
JHEP \textbf{05} (2020), 013
[arXiv:1911.12333 [hep-th]].


\bibitem{Rozali:2019day}
M.~Rozali, J.~Sully, M.~Van Raamsdonk, C.~Waddell and D.~Wakeham,
``Information radiation in BCFT models of black holes,''
JHEP \textbf{05} (2020), 004
[arXiv:1910.12836 [hep-th]].




\bibitem{Chen:2020uac}
H.~Z.~Chen, R.~C.~Myers, D.~Neuenfeld, I.~A.~Reyes and J.~Sandor,
``Quantum Extremal Islands Made Easy, Part I: Entanglement on the Brane,''
JHEP \textbf{10}, 166 (2020)
doi:10.1007/JHEP10(2020)166
[arXiv:2006.04851 [hep-th]].


\bibitem{Balasubramanian:2020hfs}
V.~Balasubramanian, A.~Kar, O.~Parrikar, G.~S\'arosi and T.~Ugajin,
``Geometric secret sharing in a model of Hawking radiation,''
JHEP \textbf{01} (2021), 177
doi:10.1007/JHEP01(2021)177
[arXiv:2003.05448 [hep-th]].


\bibitem{Akal:2020twv}
I.~Akal, Y.~Kusuki, N.~Shiba, T.~Takayanagi and Z.~Wei,
``Entanglement Entropy in a Holographic Moving Mirror and the Page Curve,''
Phys. Rev. Lett. \textbf{126} (2021) no.6, 061604
[arXiv:2011.12005 [hep-th]].


\bibitem{Kawabata:2021hac}
K.~Kawabata, T.~Nishioka, Y.~Okuyama and K.~Watanabe,
``Probing Hawking radiation through capacity of entanglement,''
[arXiv:2102.02425 [hep-th]].


\bibitem{Ishibashi:1988kg}
N.~Ishibashi,
``The Boundary and Crosscap States in Conformal Field Theories,''
Mod. Phys. Lett. A \textbf{4} (1989), 251.



\bibitem{Calabrese:2005in}
P.~Calabrese and J.~L.~Cardy,
``Evolution of entanglement entropy in one-dimensional systems,''
J. Stat. Mech. \textbf{0504} (2005), P04010
[arXiv:cond-mat/0503393 [cond-mat]].




\bibitem{Hellerman:2009bu}
S.~Hellerman,
``A Universal Inequality for CFT and Quantum Gravity,''
JHEP \textbf{08} (2011), 130
[arXiv:0902.2790 [hep-th]].

\bibitem{FZZb}
V.~Fateev, A.~B.~Zamolodchikov and A.~B.~Zamolodchikov,
``Boundary Liouville field theory. 1. Boundary state and boundary two point function,''
[arXiv:hep-th/0001012 [hep-th]].

\bibitem{Tb}
J.~Teschner,
``Remarks on Liouville theory with boundary,''
PoS \textbf{tmr2000} (2000), 041
[arXiv:hep-th/0009138 [hep-th]].

\bibitem{ZZb}
A.~B.~Zamolodchikov and A.~B.~Zamolodchikov,
``Liouville field theory on a pseudosphere,''
[arXiv:hep-th/0101152 [hep-th]].




\bibitem{Sully:2020pza}
J.~Sully, M.~Van Raamsdonk and D.~Wakeham,
``BCFT entanglement entropy at large central charge and the black hole interior,''
[arXiv:2004.13088 [hep-th]].


\bibitem{Affleck:1991tk}
I.~Affleck and A.~W.~W.~Ludwig,
``Universal noninteger 'ground state degeneracy' in critical quantum systems,''
Phys. Rev. Lett. \textbf{67} (1991), 161-164



\bibitem{Maldacena:2004rf}
J.~M.~Maldacena and L.~Maoz,
``Wormholes in AdS,''
JHEP \textbf{02} (2004), 053
[arXiv:hep-th/0401024 [hep-th]].

\bibitem{Gao:2000ga}
S.~Gao and R.~M.~Wald,
``Theorems on gravitational time delay and related issues,''
Class. Quant. Grav. \textbf{17} (2000), 4999-5008
[arXiv:gr-qc/0007021 [gr-qc]].

\bibitem{Galloway:1999br}
G.~J.~Galloway, K.~Schleich, D.~Witt and E.~Woolgar,
``The AdS / CFT correspondence conjecture and topological censorship,''
Phys. Lett. B \textbf{505} (2001), 255-262
[arXiv:hep-th/9912119 [hep-th]].

\bibitem{Kelly:2014mra}
W.~R.~Kelly and A.~C.~Wall,
``Holographic proof of the averaged null energy condition,''
Phys. Rev. D \textbf{90} (2014) no.10, 106003
[erratum: Phys. Rev. D \textbf{91} (2015) no.6, 069902]
[arXiv:1408.3566 [gr-qc]].


\bibitem{Faulkner:2016mzt}
T.~Faulkner, R.~G.~Leigh, O.~Parrikar and H.~Wang,
``Modular Hamiltonians for Deformed Half-Spaces and the Averaged Null Energy Condition,''
JHEP \textbf{09} (2016), 038
[arXiv:1605.08072 [hep-th]].

\bibitem{Hartman:2016lgu}
T.~Hartman, S.~Kundu and A.~Tajdini,
``Averaged Null Energy Condition from Causality,''
JHEP \textbf{07} (2017), 066
[arXiv:1610.05308 [hep-th]].


\bibitem{Iizuka:2019ezn}
N.~Iizuka, A.~Ishibashi and K.~Maeda,
``Conformally invariant averaged null energy condition from AdS/CFT,''
JHEP \textbf{03} (2020), 161
[arXiv:1911.02654 [hep-th]].



\end{thebibliography}
\end{document}